\documentclass[12pt]{article}

\usepackage{amsfonts}
\usepackage{amssymb}
\usepackage{amsmath}
\usepackage{latexsym}

\def\bq{\begin{quote}}
\def\eq{\end{quote}}

\newcommand{\non}{\nonumber}
\newcommand{\bea}{\begin{eqnarray}}
\newcommand{\eea}{\end{eqnarray}}
\newcommand{\ba}{\begin{array}}
\newcommand{\ea}{\end{array}}
\newcommand{\al}{\alpha}
\newcommand{\bet}{\beta}
\newcommand{\pa}{\partial}

\newcommand{\si}{\sigma}
\newcommand{\la}{\lambda}

\newcommand{\om}{\omega}

\newcommand{\De}{\Delta}

\newcommand{\tsi}{\tilde \si}

\newcommand{\rar}{\rightarrow}

\allowdisplaybreaks
\newcounter{mycount}

\begin{document}

\begin{titlepage}
\vspace{-1mm}
\begin{flushright}
\end{flushright}
\vspace{1mm}
\begin{center}{\bf\Large\sf Quantum Many--Body Problems and
Perturbation Theory}
\end{center}

\vskip 4mm

\begin{center}
 {\bf Alexander
V.~Turbiner{\normalsize \footnote{turbiner@lyre.th.u-psud.fr,\
turbiner@nuclecu.unam.mx}${}^{,} $\footnote{On leave of absence
from the Institute for Theoretical and Experimental Physics, \\
\indent \hspace{5pt} Moscow 117259, Russia.}}}\\ {\em Laboratoire
de Physique Theorique, Universit\'e Paris-Sud, France and
\\Instituto de Ciencias Nucleares, UNAM, A.P. 70-543, 04510 M\'exico}
\\[8mm]
{\bf\large Abstract}
\end{center}
\vskip 4mm

\begin{quote}
We show that the existence of algebraic forms of exactly-solvable
$A-B-C-D$ and $G_2, F_4$ Olshanetsky-Perelomov Hamiltonians allow
to develop the {\it algebraic} perturbation theory, where
corrections are computed by pure algebraic means. A classification
of perturbations leading to such a perturbation theory based on
representation theory of Lie algebras is given. In particular,
this scheme admits an explicit study of anharmonic many-body
problems. Some examples are presented.
\end{quote}

\vskip 1 cm

\begin{center}
{\it Based on invited talks at XXIII Conference `Group-Theoretical
Methods in Physics',  31.7 - 5.8.2000, Dubna, Russia\\ and\\ at
Workshop `Calogero model: thirty years after', 24-27.5.2001, Rome,
Italy}
\end{center}

\end{titlepage}
\vskip 12mm

\begin{center}
{\em INTRODUCTION}
\end{center}

Quantum integrable and exactly-solvable many-body problems
originated from projection method \cite{Olshanetsky:1977} (see
also \cite{Morozov:1990}) and/or the Hamiltonian reduction method
\cite{Kazhdan:1978} serve as a source of inspiration for many
years. The goal of this talk is to explore one more feature of
these problems -- they can be used as zero-approximation or
non-perturbed problem in order to develop constructive
perturbation theory.

We begin from some preliminary knowledge which is necessary to
enter to the subject. Take an infinite set of linear functional
spaces ${\cal V}_n,\quad n=0,1\ldots$. If they can be ordered
\[
{\cal V}_0 \subset  {\cal V}_1 \subset {\cal V}_2 \subset \ldots
 \subset  {\cal V}_n  \subset \ldots {\cal V}\ ,
\]
then such a construction is called {\em infinite flag
(filtration)} ${\cal V}$. A flag is {\it classical}, if $\dim
V_{n+1} = \mbox{dim} V_{n}+1$, otherwise it is {\it
non-classical}. If an operator $T$ such that
\[
T : {\cal V}_n \mapsto {\cal V}_n , \quad n=0,1,2, \ldots \ ,
\]
then it implies $T$ preserves the flag ${\cal V}$.

{\bf General Definition } \cite{Turbiner:1994}

An operator $T$ which preserves an infinite flag of
finite-dimensional spaces $\{ {\cal V}_k \}_{k \in \mathbb{N}}$
(namely, each space ${\cal V}_k$ is invariant to the action of
$T$) is called {\it exactly-solvable operator with flag} $\{ {\cal
V}_k \}_{k \in \mathbb{N}}$.

{\it Equivalence}

Any two functional spaces   ${\cal V}_n$ are equivalent if they
can be transformed one into another by multiplication on a
function and/or by a change of variables.

{\bf Restriction:}

{\em we study linear spaces (and flags) of polynomials only
 (and equivalent to polynomials).}

Let us consider a linear space of polynomials in ${\bf C}^d({\bf
R}^d)$
\begin{equation}
 {\cal P}^{(f)}_n = \langle\  {x_1}^{p_1}{x_2}^{p_2} \ldots
 {x_d}^{p_d} \vert\  0 \le  \sum\  \al_i p_i \le n\ \rangle \ ,
 \ n=0,1,2,\ldots \ ,
\end{equation}
where $\al_i$ are positive integers. We define the vector
\begin{equation}
 \vec{f} \ =\ (\al_1,\ldots, \al_d)\ ,
\end{equation}
which is called characteristic vector. Now one can build a flag
\begin{equation}
 {\cal P}_0^{(f)} \subset  {\cal P}_1^{(f)} \dots \subset {\cal
 P}_n^{(f)} \subset \dots \ ,
\end{equation}
which is called ${\cal P}^{(f)}$. Vector
\begin{equation}
 \vec{f_0} \ =\ (\underbrace{1,1, \ldots,1}_{d})\ ,
\end{equation}
defines the so-called {\it basic flag} ${\cal P}^{(f_0)}$ in ${\bf
C}^d ({\bf R}^d)$.

Let us consider the $gl_{d+1}$-algebra realized by
\begin{eqnarray}
\label{gld}
 {\cal J}_i^- &=& \frac{\pa}{\pa x_i},\qquad \quad
i=1,2\ldots d \ , \non \\
 {\cal J}_{ij}^0 &=&
x_i \frac{\pa}{\pa x_j}, \qquad i,j=1,2\ldots d \ ,
 \non \\
{\cal J}^0 &=& \sum_{i=1}^{d} x_i\frac{\pa}{\pa x_i}-n\, ,
 \non\\
 {\cal J}_i^+ &=& x_i {\cal J}^0 =
x_i\, \left( \sum_{j=1}^{d} x_j\frac{\pa}{\pa x_j}-n \right),
\quad i=1,2\ldots d \ ,
\end{eqnarray}
where $n \in \mathbf{C}$. If $n$ is non-negative integer, this
algebra has finite-dimensional representation and its linear space
(finite-dimensional representation space) coincides to ${\cal
P}_n^{(f_0)}$. Therefore, these finite-dimensional representation
spaces as function of $n$ being properly ordered form flag ${\cal
P}^{(f_0)}$. It is obvious that the generators ${\cal J}_i^{-,0},
{\cal J}_{ij}^0$, which span maximal affine subalgebra $b \subset
gl_{d+1}$, and their non-linear combinations preserve the flag
${\cal P}^{(f_0)}$.

{\bf Definition:}

{\it The operator $h$ is called {\em algebraic}, if it preserves a
flag of polynomials.}

\noindent It is rather obvious that algebraic operator is
characterized by polynomial coefficients, $\sum \mbox{Pol}_n \cdot
\pa^n$. It can be proven

\vskip .3cm {\bf THEOREM}
\begin{quote}
{\it Linear differential operator $h$ preserves the flag  ${\cal
P}^{(f_0)}$ iff $h= P({\cal J}(b\subset gl_{d+1}^{(*)})) $, where
$P$ is a polynomial in generators of the maximal affine subalgebra
$b$ of the algebra $gl_{d+1}$ taken in realization (\ref{gld}).}
\end{quote}

\vskip .5cm

In particular, if the second order differential operator $h$
preserves the flag  ${\cal P}^{(f_0)}$, it should have a form
\[
h = P_2^{(ij)}(x) \pa_{i}\pa_{j} + P_1^{(i)}(x) \pa_i
\]
where $P_2^{(ij)}(x)$ and $ P_1^{(i)}(x)$ are the second and first
degree polynomials in coordinates $x$'s. It is well-known {\it
hypergeometrical} operator.

\setcounter{equation}{0}

\section{\em ALGEBRAIC FORMS OF OLSHANETSKY-PERELOMOV
HAMILTONIANS}

\smallskip

In this Section we present the algebraic form of the $A_n, BC_n,
G_2, F_4$ Olshanetsky-Perelomov Hamiltonians
\cite{Olshanetsky:1977,Olshanetsky:1983}. All of them will be
obtained by the same procedure: (i) gauge rotation of the
Hamiltonian with ground state eigenfunction and (ii) a change of
variables to new variables which code symmetries of the problem.
$E_0$ is the ground state energy.

\begin{itemize}
\item {Calogero Model ($A_{N-1}$ Rational model)}\cite{Ruehl:1995}

Hamiltonian:
\[
{\cal H}_{\rm Cal} = \frac{1}{2} \sum_{i=1}^{N} \left(
-\frac{\pa^2}{\pa {x_i}^2} + \om^2 {x_i}^2 \right) + g
\sum_{i>j}^{N} \frac{1}{({x_i}-x_{j})^2}
\]
Ground state:
\begin{equation}
 \Psi_{0}^{(c)}(x) =\prod_{i<j}|x_{i}-x_{j}|^{\nu}
 e^{-\frac{\om}{2}\sum{x_i^{2}}} \ , \quad   g=\nu(\nu-1)\ .
\end{equation}
 Here
\[
 h_{{\rm Cal}} = 2(\Psi_{0}^{(c)})^{-1}\, ({\cal H}_{\rm Cal}-E_0)\,
 \Psi_{0}^{(c)}
\]
New variables:
\begin{eqnarray}
\label{v-Cal}
 Y=\sum x_i\ ,\ y_i=x_i - \frac{1}{N} Y\ ,\ i=1,\ldots , N\ ,\non \\
 (x_1,x_2,\ldots x_N) \rightarrow \big( Y, \tau_n (x)=\si_n(y(x))|
 \ n=(2 \div N) \, \big) \ ,
\end{eqnarray}
where
\[
 \si_{k}(x) = \sum_{i_{1} < i_{2} < \ldots < i_{k}} x_{i_{1}}x_{i_{2}}
 \ldots x_{i_{k}} \ ,
\]
are elementary symmetric polynomials. Finally, the gauge rotated
Calogero Hamiltonian (after separation cms)
\begin{equation}
\label{Cal}
 {h}_{\rm Cal}= {\cal A}_{ij}(\tau) \frac{\pa^2}{\pa
 {\tau_i} \pa {\tau_j} } + {\cal B}_i(\tau) \frac{\pa}{\pa \tau_i}
 \ ,
\end{equation}
 with
\begin{eqnarray}
\hspace{-10pt}{\cal A}_{ij}\hspace{-10pt} &=& \hspace{-5pt}
\frac{(N-i+1)(j-1)}{N}\,\tau_{i-1}\,\tau_{j-1} + \hspace{-20pt}
\sum_{{l\geq}{\max (1,j-i)}} \hspace{-15pt}
(j-i-2l)\,\tau_{i+l-1}\,\tau_{j-l-1} \ , \non \\[10pt]
 {\cal B}_i \hspace{-10pt}  &=& \hspace{-5pt}
 - (\frac{1}{N}+\nu){(N-i+2)(N-i+1)}\,\tau_{i-2} +2\om \,i\,
\tau_i\ .  \non
\end{eqnarray}

\item  Sutherland model ($A_{N-1}$ Trigonometric model)
\cite{Ruehl:1995}

Hamiltonian
\[
{\cal H}_{\rm Suth} =
 -\frac{1}{2}\sum_{k=1}^{N}\frac{\partial^{2}}{\partial x_{k}^{2}}
 + \frac{g}{4}\sum_{k<l}\frac{1}{\sin^{2}(\frac{1}{2}(x_{k} -
 x_{l}))}\ .
\]

Ground state
\begin{equation}
\Psi_{0}^{(s)}(x) =\prod_{i<j}
 \sin^\nu\left(\frac{1}{2}(x_{i}-x_{j})\right) \ , \
 g=\nu(\nu-1)\ .
\end{equation}
\[
h_{{\rm Suth}} = -2(\Psi_{0}^{(s)})^{-1}\, ({\cal H}_{\rm
Suth}-E_0)\, \Psi_{0}^{(s)} .
\]
New variables
\begin{equation}
\label{v-Suth}
 (x_1,x_2,\ldots x_N) \rar \big(e^{iY},\eta_n(x)=\si_n(e^{i y(x)})|
 \ {\scriptstyle n=[1 \div (N-1)]}\big) \ ,
\end{equation}
 where $y$'s are given (1.2).

Finally, the gauge rotated Sutherland Hamiltonian (after
separation cms)
\begin{equation}
\label{Suth}
 {h}_{\rm Suth} = {\cal A}_{ij}(\eta)
 \frac{\partial^2}{\pa {\eta_i} \pa {\eta_j} } + {\cal B}_i(\eta)
 \frac{\pa}{\pa \eta_i}\ ,
\end{equation}
 with
\begin{eqnarray}
{\cal A}_{ij}\hspace{-10pt} &=& \hspace{-5pt}
\frac{(N-i)\,j}{N}\,\eta_{i}\,\eta_{j} + \hspace{-10pt}
\sum_{{l\geq}{\max (1,j-i)}} \hspace{-15pt}
(j-i-2l)\,\eta_{i+l}\,\eta_{j-l}\ ,
  \non \\ {\cal B}_i
\hspace{-10pt}  &=& \hspace{-5pt}
  (\frac{1}{N}+\nu)\,i\,(N-i)\,\eta_{i} \ .
   \non
\end{eqnarray}

\item{$BC_N$ --Rational model} \cite{Brink:1997}

Hamiltonian
\begin{eqnarray}
\hspace{-10pt}{\cal H}^{(r)}_{BC_N} \hspace{-10pt}&=&\hspace{-5pt}
-\frac{1}{2}\sum_{i=1}^{N}\left( \frac{\pa^2}{\partial {x_i}^2} -
\omega^{2}x_{i}^{2}\right) + g\sum_{i<j}\left[
\frac{1}{(x_{i}-x_{j})^{2}} + \frac{1}{(x_{i}+x_{j})^{2}} \right]
\non \\ &&+ \ \frac{g_{2}}{2}\sum_{i=1}^{N}\frac{1}{x_{i}^{2}} \ .
\non
\end{eqnarray}
Ground state
\[
 \Psi_{0} = \left[
\prod_{i<j}|x_{i}-x_{j}|^{\nu}|x_{i}+x_{j}|^{\nu}
\prod_{i=1}^{N}|x_{i}|^{\nu_{2}} \right]
e^{-\frac{\om}{2}\sum_{i=1}^{N}x_{i}^{2}} \ ,
\]
\begin{equation}
 g=\nu(\nu-1)\ ,\ g_2=\nu_2(\nu_2-1) \ .
\end{equation}
Here
\[
h_{{BC}_N}^{(r)} = -2(\Psi_{0})^{-1}\, ({\cal
H}_{BC_N}^{(r)}-E_0)\, \Psi_{0} \ .
\]
 New variables
\begin{equation}
\label{v-BCr}
 (x_1,x_2,\ldots x_N) \rightarrow \big(\si_k(x^2)| \
 k=(1 \div N)\big) \ .
\end{equation}
 Finally, the gauge rotated $BC_N$ rational Hamiltonian
\begin{equation}
\label{HBCr}
 {h}^{(r)}_{BC_N}= {\cal A}_{ij}(\si)
 \frac{\pa^2}{\pa {\si_i} \pa {\si_j} } + {\cal B}_i(\si)
 \frac{\pa}{\pa \si_i}\ ,
\end{equation}
 with
\begin{eqnarray}
{\cal A}_{ij} &=& 4\, \sum_{l\ge 0} (2l+1+j-i)\,
\si_{i-l-1}\,\si_{j+l}\ ,
 \non \\[10pt]
{\cal B}_i &=& 2\, \left[ 1+\nu_2 + 2\nu(N-i)\right]\left[
N-i+1\right]\, \si_{i-1} -4\,\om\,i\,\si_i \ .
 \non
\end{eqnarray}

\end{itemize}

\begin{itemize}
\item{$BC_N$ --Trigonometric model} \cite{Brink:1997}

Hamiltonian:
\begin{eqnarray}
 \hspace{-20pt}{\cal H}^{(t)}_{BC_N} \hspace{-10pt}&=& \hspace{-10pt}
-\frac{1}{2}\sum_{i=1}^{N} \frac{\pa^2}{\pa {x_i}^2} \!+
\frac{g}{4} \sum_{i<j} \left[
\frac{1}{\sin^2\!\left(\frac{1}{2}(x_{i}-x_{j})\right)} +
\frac{1}{\sin^2\!\left(\frac{1}{2}(x_{i}+x_{j})\right)} \right]
\non \\ && + \ \frac{g_{2}}{4} \sum_{i=1}^{N} \frac{1}{\sin^2\!
x_{i} } + \ \frac{g_{3}}{4} \sum_{i=1}^{N} \frac{1}{\sin^2\!
\frac{x_{i}}{2}}\ . \non
\end{eqnarray}
Ground state:
\[
 \Psi_{0} \ =\ \left[ {\prod_{i<j}} |\sin( {\frac{x_{i}-x_{j}}{2}
})|^{\nu} |\sin( { \frac{x_{i}+x_{j}}{2} })|^{\nu}
{\prod_{i=1}^{N}} |\sin({ x_{i}})|^{\nu_{2}} |\sin({
\frac{x_{i}}{2}})|^{\nu_{3}} \right] \ ,
\]
\begin{equation}
g = \nu(\nu - 1)\ ,\ g_{2} = \nu_{2}(\nu_{2} - 1)\ , \ g_{3} =
\nu_{3}(\nu_{3} + 2\nu_{2} - 1)\ .
\end{equation}
 Here
\[
h_{{BC}_N}^{(t)} = -2(\Psi_{0})^{-1}\, ({\cal
H}_{BC_N}^{(t)}-E_0)\, \Psi_{0} \ .
\]
New variables
\begin{equation}
\label{v-BCt}
 (x_1,x_2,\ldots x_N) \rightarrow \big(
\hat\si_k(x)=\si_k(\cos x)| \ { k=(1 \div N)}\big) \ .
\end{equation}
Finally, the gauge rotated $BC_N$ trigonometric Hamiltonian
\begin{equation}
\label{HBCt}
 {h}^{(t)}_{BC_N} = {\cal A}_{ij}({\hat\si}) \frac{\pa^2}{\pa
 {{\hat\si}_i} \pa {{\hat\si}_j} } + {\cal B}_i({\hat\si})
 \frac{\pa}{\pa {\hat\si}_i} \ ,
\end{equation}
 with
\begin{eqnarray}
\hspace{-10pt}{\cal A}_{ij} \hspace{-10pt}&=&\hspace{-10pt}
N\,{\hat\si}_{i-1}\,{\hat\si}_{j-1} - \hspace{-5pt} \, \sum_{l\ge
0} \Big[
  (i-l)   \,{\hat\sigma}_{i-l}  \,{\hat\sigma}_{j+l}
+ (l+j-1) \,{\hat\sigma}_{i-l-1}\,{\hat\sigma}_{j+l-1}
 \non\\
&& - (i-2-l) \,{\hat\si}_{i-2-l}\,{\hat\si}_{j+l} - (l+j+1)
\,{\hat\si}_{i-l-1}\,{\hat\si}_{j+l+1} \Big]\ ,
 \non\\[5pt]
{\cal B}_i  \hspace{-10pt}&=&\hspace{-10pt}
\frac{\nu_3}{2}(i-N-1)\,{\hat\si}_{j-1}
-
\Big[\nu_2 + \frac{\nu_3}{2}+1+\nu(2N-i-1)\,i\,{\hat\si}_{i}
 \non\\
&&  \hspace{110pt} - \nu(N-i+1)(N-i+2){\hat\si}_{i-2} \Big]\ .
 \non
\end{eqnarray}

\item{$G_2$ --Rational model} \cite{Rosenbaum:1998}

Hamiltonian:
\begin{eqnarray}
{\cal H}_{G_2}^{(r)} & = & -\frac{1}{2}\sum_{i=1}^{3}\left(
\frac{\pa^2}{\pa {x_i}^2} - \om^{2}x_{i}^{2}\right) + g\sum_{i<j}
\frac{1}{(x_{i}-x_{j})^{2}} \non \\ && + \ g_1
\sum_{i<j}{\frac{1}{(x_{k} + x_{l}-2x_{m})^2}} \ .
 \non
\end{eqnarray}
Ground state
\begin{eqnarray}
{ \Psi_{0} \!=\hspace{-5pt} \prod_{i<j}^3|x_i-x_j|^{\nu}
\prod_{\mbox{$\begin{array}{c}\scriptstyle i<j \\ \scriptstyle
i,j\neq k\end{array}$} } |x_i+x_j-2x_k|^{\mu}\,e^{-\frac{1}{2}\om
\sum x_i^2}}\ , \non
\\
g = \nu(\nu - 1) > -\frac{1}{4} , \ g_1=3\mu (\mu -1) >
-\frac{3}{4} \ .
\end{eqnarray}
Here
\[
h_{{G}_2}^{(r)} = -2(\Psi_{0})^{-1}\, ({\cal H}_{G_2}^{(r)}-E_0)\,
\Psi_{0} \ .
\]
New variables
\[
Y=\sum x_i\ ,\ y_i=x_i - \frac{1}{3} Y\ ,\ i=1,2,3\ ,
\]
and
\begin{equation}
\label{v-G2r}
 (x_1,x_2,x_3) \rar \big(Y,\ \la_1(y),\ \la_2(y)\big) \ ,
\end{equation}
\[
 \la_1=-y_1^2-y_2^2-y_1 y_2\ ,\ \la_2=[y_1 y_2(y_1+y_2)]^2 \ .
\]
 Finally, the gauge rotated $G_2$ rational Hamiltonian (after
 separation cms)
\[
h_{\rm G_2}^{(r)}  =  -2\la_1\pa^2_{\la_1\la_1}
        -12\la_2\pa^2_{\la_1\la_2}
        +\frac{8}{3}\la_1^2\la_2\pa^2_{\la_2\la_2}
\]
\begin{equation}
\label{HG2r}
   -\big\{4\om\la_1+2[1+3(\mu+\nu)]\big\}\pa_{\la_1}
   -\bigl( 12\om\la_2-\frac{4}{3}\la_1^2\bigr)\pa_{\la_2}\ .
\end{equation}

\item{$G_2$ --Trigonometric model} \cite{Rosenbaum:1998}

Hamiltonian
\[
{\cal H}_{\rm G_2}^{(t)} =
 -\frac{1}{2}\sum_{k=1}^{3}\frac{\pa^{2}}{\pa x_{k}^{2}}
 + \frac{g\al^2}{4}\sum_{k<l}^{3}\frac{1}{\sin^{2}(\frac{\al}{2}(x_{k} - x_{l}))}
\]
\[
 + \frac{g_1 \al^2}{4}\sum_{ k<l, \ \ k,l \neq m}^{3}
 \frac{1}{\sin^{2}(\frac{\al}{2}(x_{k} + x_{l}-2x_{m}))}
\]

Ground state
\begin{eqnarray}
{\Psi_{0} \!=\hspace{-5pt} \prod_{i<j}^3 |\sin\frac{\al}
{2}(x_{i}-x_{j})|^{\nu} \prod_{\mbox{$\begin{array}{c}
\scriptstyle k<l \\ \scriptstyle k,l \neq
m\end{array}$}}^3|\sin\frac{\al} {2}(x_{i}+x_{j}-2x_{k})|^{\mu}}\
. \non
\\
g = \nu(\nu - 1) > -\frac{1}{4} , \ g_1=3\mu (\mu -1) >
-\frac{3}{4} \ ,
\end{eqnarray}
Here
\[
h_{{G}_2}^{(t)} = -2(\Psi_{0})^{-1}\, ({\cal H}_{G_2}^{(t)}-E_0)\,
\Psi_{0} \ .
\]
New variables
\[
Y=\sum x_i\ ,\ y_1=x_1 - x_2,\ y_2=x_2 - x_3,\ y_3=x_3 - x_1,
\]
\begin{equation}
\label{v-G2t}
 (x_1,x_2,x_3) \rar \big(Y,\ \tsi_1,\ \tsi_2\big) \ ,
\end{equation}
\[
\tsi_1  = \frac{1}{\al^2}\biggl[\cos\al(y_1-y_2)+\cos\al(
y_2-y_3)+\cos\al( y_3-y_1)-3\biggr]\ , \non
\]
\[
 \tsi_2  =  \frac{4}{\al^6} \biggl[ \sin\al( y_1-y_2)+\sin\al(
 y_2-y_3)+\sin\al( y_3-y_1)\biggr]^2 \ .
\]

Finally, the gauge rotated $G_2$ trigonometric Hamiltonian (after
separation cms)
\[
h_{G_2}^{(t)} \ = \
-(2\tsi_1+\frac{\al^2}{2}\tsi_1^2-\frac{\al^4}{24}\tsi_2)\pa_{\tsi_1\tsi_1}^2-
(12+\frac{8\al^2}{3}\tsi_1)\tsi_2\pa_{\tsi_1\tsi_2}^2\ +
\]
\[
+(\frac{8}{3}\tsi_1^2\tsi_2-2\al^2\tsi_2^2)\pa_{\tsi_2\tsi_2}^2 -
\bigl\{2[1+3(\mu +2\nu)]+\frac{2}{3}(1 + 3\mu +
       4 \nu)\al^2\tsi_1\bigr\}\pa_{\tsi_1}\ +
\]
\begin{equation}
\label{HG2t}
 \bigg\{\frac{4}{3}(1+4\nu)\tsi_1^2-[\frac{7}{3}
 +4(\mu +\nu )]\al^2\tsi_2\bigg\}\pa_{\tsi_2}\ .
\end{equation}

\item{$F_4$-Rational model} \cite{Boreskov:2001}

Hamiltonian
\begin{align}
 {\cal H}_{\rm F_4}^{(r)} = &\ \frac{1}{2}\
\sum_{i=1}^{4} \left( -\pa_{x_i}^2 + 4\om^2 x_i^2 \right) + 2g
\sum_{j>i} \left( \frac{1}{(x_i-x_j)^2} + \frac{1}{(x_i+x_j)^2}
\right)\non \\ &+ 2g_1 \sum_{i=1}^{4}\frac{1}{{x_i}^2} + 8g_1
\sum_{ \nu's=0,1} \frac{1}{ \left[ x_1 + (-1)^{\nu_2}x_2+
(-1)^{\nu_3}x_3+ (-1)^{\nu_4}x_4 \right]^2}\non\ .
\end{align}

Ground state
\begin{align}
 & \Psi_{0}^{({\rm r})}(x) = \left(\De_-\De_+\right)^{\nu}
 \left(\De_0 \De \right)^{\mu}
\exp\left(-\om\sum_{i=1}^{4} {x_i}^2\right)\ ,\non
\end{align}
\begin{equation}
g=\nu(\nu-1)/2\ ,\ g_1 =\mu(\mu-1)\, ,
\end{equation}
 with
\begin{align}
 \De_{\pm} &= \prod_{j<i}^4 (x_i\pm x_j) \ , \non\\
 \De_{0} &= \prod_{i=1}^4 x_i \ , \non \\
 \De &= \prod_{\nu's=0,1}
 \left[x_1 + (-1)^{\nu_2}x_2 +
 (-1)^{\nu_3}x_3+ (-1)^{\nu_4}x_4 \right]\ . \non
\end{align}

New variables
\begin{equation}
\label{v-F4r}
 (x_1,x_2,x_3,x_4) \rar \big(t_1,t_3,t_4,t_6 \big)\ ,
\end{equation}
 where
\[
 t_1 = \si_1\ ,
\]
\[
 t_3 = \si_3-\frac {1}{6} \si_1\, \si_2 \ ,
\]
\[
 t_4 = \si_4-\frac {1}{4}\, \si_1\, \si_3 +
 \frac{1}{12}\, \si_2^{2}\ ,
\]
\[
 t_{6} = \si_4 \, \si_2 -{\frac {1}{36}}\, \si_2^{3} -
 \frac {3}{8}\, \si_3^{2}+\frac {1}{8}\, \si_1\,
 \si_2 \, \si_3 - \frac {3}{8}\, \si_1^2\, \si_4 \ .
\]
and $\si_a= \si_a(x^2)$.

Finally, gauge rotated $F_4$ rational Hamiltonian
\begin{equation}
\label{HF4r}
 h_{\rm F_4}^{(r)}
= \, {\cal A}_{ab} \frac{\pa^2} {\pa
 t_a\pa t_b}\ +\,  \left({\cal B}_a + {\cal C}_a \right)\frac{\pa\ } {\pa t_a} \ ,
\end{equation}
with
\begin{alignat}{2}
 & {\cal A}_{11} = 4\, t_{1}  &\qquad
 & {\cal A}_{13} = 12\, t_{3} \  ,\non\\[3pt]
 & {\cal A}_{14} = 16\, t_{4}\ ,  &
 & {\cal A}_{16} = 24\, t_{6}\ ,\non\\[3pt]
 & {\cal A}_{33} = - { \frac {2}{3}}
\,{ t_{1}}^{2}\,{ t_{3}} + { \frac {20}{3}} \,{ t_{1}}\,{ t_{4}} \
, &
 & {\cal A}_{34} = - {\frac {4}{3}}
\, t_{1}^{2}\, t_{4} + 8\, t_{6}\ , \non\\
 & {\cal A}_{36} = 16\,{ t_{4}}^{2} -
2\, t_{1}^{2}\, t_{6}\ ,  &
 & {\cal A}_{44} = - 4\, t_{3}\, t_{4} -
2\, t_{1}\, t_{6}\ , \non\\[3pt]
 & {\cal A}_{46} = - 4\, t_{1}\, t_{4}^{2} -
6\, t_{3}\, t_{6}\ , &
 & {\cal A}_{66} = - 12\, t_{3}\, t_{4}^{2} -
 6\, t_{1}\, t_{4}\, t_{6} \ ,\non
\end{alignat}
\[
  {\cal A}_{b\,a} =\ {\cal A}_{a\,b} \ ,
\]
\begin{alignat}{2}
 & {\cal B}_{1} = 8\ ,&\qquad
 & {\cal B}_{3} = - \, t_{1}^{2} \ ,\non \\
 & {\cal B}_{4} = - 4\, t_{3} \ ,&
 & {\cal B}_{6} = - 8\, t_{1} t_{4} \ . \non
\end{alignat}
\begin{alignat}{2}
 &{\cal C}_{1} =  48(\nu +\mu)-4\om t_1 \ , &\quad&
  {\cal C}_{3} = -2(2\nu+\mu) t_1^2 -12\om t_3\ ,\non\\
 &{\cal C}_{4} = -12 \nu  t_3 - 16 \om t_4 \ , &&
  {\cal C}_{6} = -12 \nu  t_1 t_4 - 24 \om t_6 \ .\non
\end{alignat}

\item{$F_4$-Trigonometric model} \cite{Boreskov:2001}

Hamiltonian
\begin{equation}
  {\cal H}_{\rm F_4}^{(t)}(x) \ =\ -\frac{1}{2} \sum_{i=1}^{4}
 \partial_{x_i}^2 + 2g V_1(x,\beta) +
 \frac{g_1}{2} V_2(x,2\beta) \ ,
\end{equation}
where $g=\nu (\nu-1)/2$, $g_1=\mu (\mu-1)$, and
\begin{align}
 V_1(x,\beta) = &\ \beta^2 \sum_{j>i} \left(
\frac{1}{\sin^2 \beta(x_i-x_j)} + \frac{1}{\sin^2 \beta(x_i+x_j)}
\right)\ ,\non \\
 V_2(x,2\beta) = &\ 4\beta^2 \sum_{i=1}^{4}\frac{1}{\sin^2 2\beta{x_i}}  \non\\
+\ & 4\beta^2 \sum_{ \nu's=0,1}^4 \frac{1}{\sin^2 \beta \left[ x_1
+ (-1)^{\nu_2}x_2+ (-1)^{\nu_3}x_3+ (-1)^{\nu_4}x_4 \right]}\
.\non
\end{align}
Ground state
\begin{align}
\label{e5.1}
 \Psi_0^{(t)} (x,\beta) =
 \left( \De_+(x,\bet) \De_-(x,\bet) \right)^{\nu}
 \left(\Delta_0 (x,2\bet) \Delta (x,2\bet) \right)^{\mu} \ ,
\end{align}
where
\begin{align}
 \De_{\pm}(x,\beta) &=
 \beta^{-6}\prod_{j<i}\sin\bet(x_i\pm x_j) \ , \non\\
 \De_{0}(x,2\beta) &=
 \beta^{-4}\prod_{i}\sin 2\beta x_i \ , \non\\
 \De (x,2\beta) &= \beta^{-8}\prod_{\nu's=0,1}\sin \beta\left[x_1 +
 (-1)^{\nu_2}x_2 + (-1)^{\nu_3}x_3+ (-1)^{\nu_4}x_4 \right] \
 .\non
\end{align}
Here
\begin{align}
 h_{\rm F_4}^{(t)} \ =\ -2\big(\Psi_{0}^{(t)}(x)\big)^{-1}({\cal
 H}_{\rm F_4}^{(t)}-E_0) \big(\Psi_{0}^{(t)}(x)\big) \ .\non
\end{align}
New variables
\begin{equation}
\label{v-F4t}
 (x_1,x_2,x_3,x_4) \rar \big(\tau_1,\tau_3,\tau_4,\tau_6 \big)\ ,
\end{equation}
 where
\begin{align}
\label{e5.4}
 \tau_{1} =& \,\si_{1}-\frac{2\bet^2}{3}\si_{2}\ , \non\\
 \tau_{3} =& \,\si_3 - \frac{1}{6}\,\si_1\,\si_2-
2\beta^2(\si_{4}-\frac{1}{36}\si_{2}^2)\ , \non\\
 \tau_{4} =& \,\si_4 - {\frac{1}{4}}\,\si_1\,\si_3+
{\frac{1}{12}}\,\si_2^{2}\ , \non\\
 \tau_{6} =& \,\si_4\,\si_2 - {\frac{1 }{36}}\,\si_2^{3}-
 {\frac{3}{8}}\,\si_3^{2}+\frac{1}{8}
\,\si_1\,\si_2\,\si_3 - \frac{3}{8}\,\si_1^{2}\,\si_4\ .
\end{align}
and $\si_a=\si_a(y^2)\ ,\ y_i = \frac{\sin(\bet x_i )}{\bet}$.

Finally, the gauge-rotated $F_4$ trigonometric Hamiltonian
\begin{align}
\label{HF4t}
 h_{\rm F_4}^{(t)}
= \, A_{ab} \frac{\pa^2}{\pa\tau_a\pa\tau_b}\ +\,
  \left(B_a + C_a\right)\frac{\pa\ } {\pa\tau_a} \ ,
 \qquad a,b = 1,3,4,6 \ ,
\end{align}
where the coefficient functions are
\begin{align}
 A_{11} &= 4\,\tau_{1}-4\beta^2{
\tau}_{1}^2-\frac{32}{3}\beta^4\tau_{3}-\frac{128}{9}\beta^6
\tau_{4}\ ,\non\\[5pt]
 A_{13} &= 12\,\tau_{3}-\frac{8}{3}\beta^2(4\tau_{1}\tau_{3}+\tau_{4})-
\frac{32}{9}\beta^4\tau_{1}\tau_{4}\ ,\non\\[5pt]
 A_{14} &=  16\,\tau_{4}
-\frac{40}{3}\beta^2\tau_{1}\tau_{4}- \frac{16}{3}\beta^4\tau_{6}
 \ ,\non \\[5pt]
 A_{16} &=   24\,\tau_{6}
- 20\beta^2\tau_{1}\tau_{6}- \frac{32}{3}\beta^4\tau_{4}^2
 \ ,\non\\[5pt]
A_{33} &=  -\frac {2}{3} \,\tau_{1}^{2}\,\tau_{3} +\frac {20}{3}
\,\tau_{1}\,\tau_{4} - \frac{8}{9}\beta^2\,(18\tau_{3}^2
+\tau_{1}^2\,\tau_{4}
 +12\tau_{6}) \ ,
 \non\\[5pt]
A_{34} &= -\frac{4}{3} \,\tau_{1}^{2}\, \tau_{4} + 8\,\tau_{6}
 -\frac{4}{3}\beta^2\,(\tau_{1}\,\tau_{6}
 +12\tau_{3}\,\tau_{4})\ ,
 \non \\[5pt]
A_{36} &= 16\,\tau_{4}^{2} - 2\,\tau_{1}^{2}\,\tau_{6}
 -\frac{8}{3}\beta^2 (9\tau_{3}\,\tau_{6}
 +\tau_{1}\,\tau_{4}^2)\ ,
 \non\\[5pt]
A_{44} &= - 4\,\tau_{3}\,\tau_{4} -
2\,\tau_{1}\,\tau_{6}-24\beta^2\tau_{4}^2 \ ,\non\\[9pt]
A_{46} &= - 4\,\tau_{1}\,\tau_{4}^{2} - 6 \,\tau_{3}\,\tau_{6}-
36\beta^2 \tau_{4}\tau_{6} \ ,\non\\[9pt]
A_{66} &= - 12\tau_3\tau_4^2 - 6 \tau_{1}\tau_{4}\tau_6 -
8\beta^2(6\tau_{6}^2 +\tau_{4}^3) \ , \non
\end{align}
\[
  A_{b\,a} \ =\ A_{a\,b} \ ,
\]
\begin{alignat}{3}
 &B_1 = 8-8\beta^2 \tau_1 \ , &\qquad&
B_3 = - \tau_{1}^{2}- \frac{56}{3}\beta^2\tau_{3}
-\frac{32}{9}\beta^4\tau_{4}\ ,\non\\
 &B_4 = - 4\,\tau_{3}-\frac{88}{3}\beta^2\tau_{4}
\  , &&
 B_6 = -8 \tau_{1}\tau_{4}- 56\beta^2 \tau_6 \ .\non
\end{alignat}
\begin{alignat}{3}
 &C_{1} = 48(\nu+\mu) -8\bet^2(5\nu+6\mu)
\tau_1 \ , &\quad&
 C_{3}=-2(2\nu+\mu)\tau_1^2-16\bet^2(3\nu+5\mu)\tau_3 \ , \non\\[3pt]
 &C_{4}= -12\nu\tau_3-24\bet^2 (3\nu+4\mu)\tau_4 \ ,&&
 C_{6} = -12\nu\tau_1\tau_4 -48\bet^2(2\nu+3\mu)\tau_6 \ .\non
\end{alignat}

\end{itemize}

\centerline{\large Remarks and Comments}

\begin{itemize}

\item
$A_N-$ and $BC_N-$ rational and trigonometric models possess {\bf
algebraic} forms; their Hamiltonians (\ref{Cal}), (\ref{Suth}),
(\ref{HBCr}), (\ref{HBCt}) preserve the {\bf same} basic flag of
polynomials ${\cal P}^{(f_0)}$.

\item
All $A_N-$ and $BC_N-$ rational and trigonometric Hamiltonians
taken in algebraic form can be written as
\[
h = P_2({\cal J}(b\subset gl_{N+1}))
\]
where $P_2$ is a polynomial of second degree in the generators
${\cal J}$ of the maximal affine subalgebra of the algebra
$gl_{N+1}$ in realization (\ref{gld}). One can state that
$gl_{N+1}$ is their {\it hidden algebra}.

\item Both rational and trigonometric $G_2$ models possess
algebraic forms; their Hamiltonians preserve the {\bf same} flag
of polynomials ${\cal P}^{(f_{G_2})}$ with $\vec{f}_{G_2}=(1,2)$;
their hidden algebras coincide and it is some
infinite-dimensional, finitely-generated algebra $g^{(2)} \subset
{\mbox{diff}}({\mathbb C}^2)$ (see \cite{Rosenbaum:1998}).

\item Both rational and trigonometric $F_4$ models possess
algebraic forms; their Hamiltonians preserve the {\bf same} flag
of polynomials ${\cal P}^{(f_{F_4})}$ with
$\vec{f}_{F_4}=(1,2,2,3)$; their hidden algebras coincide and it
is some infinite-dimensional, finitely-generated algebra $f^{(4)}
\subset {\mbox{diff}}({\mathbb C}^4)$ (see \cite{Boreskov:2001}).

\item New variables (\ref{v-Cal}), (\ref{v-Suth}), (\ref{v-BCr}),
(\ref{v-BCt}), (\ref{v-G2r}), (\ref{v-G2t}), (\ref{v-F4r}),
(\ref{v-F4t}), in which the algebraic forms occur, usually absorb
all external symmetries of model under investigation; they have a
meaning of rational and trigonometric invariants in the
corresponding root space; to the best of our knowledge they were
used for the first time to find flat space metrics (denoted by
${\cal A}$ in $A-B-C-D$ and $F_4$ examples) in rational case by
V.I.~Arnold \cite{Arnold:1976}, we will call these metrics ${\cal
A}$ the {\bf Arnold metrics}.

\item
Although the question about existence of the algebraic forms for
rational and trigonometric $E_{6,7,8}$ models was not
constructively studied yet, there are almost no doubts that they
should exist.

\end{itemize}

\setcounter{equation}{0}

\section{ \em PERTURBATION THEORY}

Existence of algebraic forms leads to a possibility to construct a
special, {\it algebraic} perturbation theory -- a type of
perturbation theory where {\bf finding corrections is an algebraic
procedure} and furthermore any correction has a form of
finite-order polynomial in coordinates.

Consider the spectral problem,
\begin{equation}
\label{pt.1} (T_0 + \lambda T_1)\phi = E \phi \ ,
\end{equation}
where $\la$ is a formal parameter, and let us develop perturbation
theory:
\begin{equation}
\label{pt.2}
 \phi = \sum \lambda^k \phi_k \ ,\ E = \sum \lambda^k
 E_k \ .
\end{equation}
Then the following theorem holds:

{\bf THEOREM}
\begin{quote}
{\it Let $T_0$ be an exactly-solvable operator with flag $\{ {\cal
 V}_k\}_{k \in \mathbb{N}}$. Let the perturbation $T_1 $ is such
 that $T_1$ is an element of space ${\cal V}_n$ from the flag
 and we look for $\phi \in {\cal V}$.
 Then the perturbation theory is algebraic:
 $\exists p(k)$ such that $k$-th correction $\phi_k \in
 {\cal V}_{p(k)}$ and hence it can be found by algebraic means.}
\end{quote}

The proof is quite straightforward and is based on analysis of the
equation for $k$th correction
\[
(T_0 - E_0) \phi_k = \sum_{i=1}^{k} E_i \phi_{k-i} - T_1
\phi_{k-1}\ .
\]
We can proceed to examples.

\noindent {\bf Example 1.} One-dimensional Anharmonic Oscillator.

\noindent It is characterized by the Hamiltonian
\begin{equation}
\label{AHO}
 {\cal H} = \underbrace{-\frac{1}{2} \frac{\pa^2}{\pa
y^2} + \omega^2\,y^2 + \frac{g}{y^2}}_{{
A_1-}\mbox{\small{Calogero model}}} + \la \, y^4 \ .
\end{equation}
 Ground state
\begin{equation}
 \psi_0 = y^\nu\, e^{-\frac{\omega}{2}\,y^2}\ ,\quad g=\nu(\nu-1)\
 \ ,\quad E_0=\om (1 + 2\nu)\ .
\end{equation}
 In new variable
\[
 \tau = y^2\ ,
\]
the gauge-rotated Hamiltonian
\[
h = \frac{1}{\om}\,\psi_0^{-1} ({\cal H}-E_0) \psi_0 \ = \non
\]
\[
\ -2\tau \partial^2_\tau +2(\tau-\mu)\partial_\tau + \lambda\tau^2
\equiv T_0 + \lambda T_1 \ ,
\]
where $ \mu \equiv \nu+1/2$. It is easy to check that
\[
T_0 : {\cal P}_n \mapsto {\cal P}_n, \quad E_0^{(n)} = 2n, \quad
n=0,1,2,\ldots,
\]
\[
T_1 = \tau^2 \in {\cal P}_{2,3,\ldots}
\]
where ${\cal P}$ is basic flag of polynomials in $\bf C$ (see
(0.1)).

\begin{description}

\item[1.]
Ground state:

Now the ground state of $T_0$ is is given by $\phi_0^{(0)}=1\ , \
E_0^{(0)}=0$ and the

\noindent {\it First correction:}

Defining equation is
\[
 -2\tau \pa^2_\tau \phi_1^{(0)} +2(\tau-\mu)\pa_\tau\phi_1^{(0)} =
E_1^{(0)} - \tau^2 \ ,
\]
with a solution
\begin{eqnarray}
 -\phi_1^{(0)} = \frac{1}{4} \tau^2 + \frac{\mu+1}{2} \tau \ ,
\\
E_1^{(0)} = \mu(\mu+1)\ .
\end{eqnarray}

\noindent {\it Second correction:}
\[
 -2\tau \pa^2_\tau \phi_2^{(0)} +2(\tau-\mu)\pa_\tau\phi_2^{(0)} =
E_2^{(0)} + E_1^{(0)} \phi_1^{(0)} - \tau^2 \phi_1^{(0)} \ ,
\]
\[
 \phi_2^{(0)}\ =\ \frac{\tau^4}{32}+
 \frac{3\mu+4}{24}\tau^3\ +\ \frac{2\mu^2+10\mu+9}{16}\tau^2\ +
 \frac{(\mu+1)(4\mu+5)}{4}\tau \ ,
\]
\[
 E_2^{(0)}\ =\ -\frac{\mu(\mu+1)(4\mu+5)}{2} \ .
\]
In general, an arbitrary correction to the ground state has a from
\[
 \phi_k^{(0)} = a_{2k} \tau^{2k}+a_{2k-1}
 \tau^{2k-1}+\ldots+a_{2k-m} \tau^{2k-m}+\ldots \ .
\]
Coefficients in front of leading terms can be found explicitly for
any excited state(!) - they are generalized Catalan numbers of a
form
\[
a_{2k-m} \sim \frac{(2k)!}{k! (k-m/2)!} \ .
\]

In standard Rayleigh-Schroedinger Perturbation Theory (RSPT) the
first energy correction $E_1^{(0)} \ =\ \langle 0 \vert T_1 \vert
0 \rangle/ \langle 0 \vert 0 \rangle$, hence
\[
E_1^{(0)} \ =\  \frac{\langle 0 \vert y^4  \vert 0
\rangle}{\langle 0 \vert 0 \rangle}\ =\ \mu(\mu+1)
\]
therefore we can find the expectation value $\langle 0 \vert y^4
\vert 0 \rangle$ {\it algebraically} (up to known normalization
factor (see e.g. \cite{Olshanetsky:1983}). A comparison of other
corrections in present perturbation theory and RSPT allows to find
algebraically transition amplitudes between different states
(correlation functions).

\item[2.]
 First Excited State: $\qquad \phi_0^{(1)}\ =\ \tau -\mu, \quad
E_0^{(1)}=2$

\noindent {\it First correction:}

Defining equation
\[
 -2\tau \pa^2_\tau \phi_1^{(1)} +2(\tau-\mu)\pa_\tau\phi_1^{(1)} -
 2\phi_1^{(1)} =
(E_1^{(1)} - \tau^2)(\tau -\mu)\ ,
\]
and the correction
\begin{equation}
-\phi_1^{(1)} = \frac{1}{4}[\tau^3 - (\mu-3)
\tau^2+2(\mu+1)(\mu-3)\tau ] \ ,
\end{equation}
\begin{equation}
 E_1^{(1)} = -(\mu+1)(\mu-3)\ .
\end{equation}

\end{description}

It is worth to note that the developed perturbation theory in
present example coincides to the so-called Dalgarno-Lewis form of
perturbation theory \cite{Dalgarno}. In fact, it was namely this
form of perturbation theory which was successfully applied by
Bender and Wu \cite{Bender} in their profound study of the problem
(\ref{AHO}) at $g=0$.

\noindent {\bf Example 2}. $(N-1)$-dimensional Anharmonic
Oscillator.

Consider the following perturbed $N$-body Calogero model
\[
{\cal H} = {\cal H}_{Cal} + \lambda\, \tau_4(x),\quad  N>4
\]
\[
\tau_4(x)= \si_{4}(y) = \sum_{i_{1},i_{2},i_{3},i_{4}}
y_{i_{1}}y_{i_{2}}y_{i_{3}}y_{i_{4}}
\]
\[
h= h_{Cal} + \lambda\,\tau_4 \equiv T_0 + \lambda\,T_1
\]
\[
T_0 : {\cal P}_n^{(N-1)}(\tau) \mapsto {\cal P}_n^{(N-1)}(\tau),
\quad n\in \mathbb{N},
\]
\[
T_1 = \tau_4 \in {\cal P}_{1,2,3,\ldots}^{(N-1)} \ ,
\]

Ground State is given by

\[
 \phi_0^{(0)}=1\ , \quad
E_0^{(1)}=0\ .
\]

\noindent {\it First correction:}

\[
 -\phi_1^{(0)} = \frac{1}{8\om}\,\tau_4 +
 \frac{1}{32\om^2}\bigg(\frac{1}{N}+\nu\bigg)(N-2)(N-3)\,\tau_2
\]
\[
E_1^{(0)}= \frac{1}{32\om^2}\bigg(\frac{1}{N}+\nu\bigg)^2
\frac{N!}{(N-4)!}
\]
Again we can find expectation value {\it algebraically} (up to
known normalization factor).
\[
E_1^{(0)} \ =\ \frac{\langle 0 \vert \tau_4(y) \vert 0
\rangle}{\langle 0 \vert 0 \rangle} \ .
\]

{\it Second correction} is of the form
\[
 \phi_2 = \al_1\,\tau_2^2+\al_2\,\tau_3^2+\al_3\,\tau_4^2+\al_4
 \,\tau_2 \tau_4 +
   \bet_1\,\tau_2+\bet_2\,\tau_4 + \bet_3\,\tau_6 \ ,
\]
where the coefficients $\al$'s and $\bet$'s can be easily
computed.

{\bf Example 3}. Perturbed 3-body Sutherland Model.

Take
\[
{\cal H} = {\cal H}_{\rm Suth}^{(3)} + \lambda\,\eta_2
\]
where ${\cal H}_{\rm Suth}^{(3)}$ is the Hamiltonian of 3-body
Sutherland model. Gauging away the ground state (1.4) and
introducing new variables
\[
\eta_2  =  \frac{1}{\alpha^2}[\cos(\al y_1)+\cos(\al
y_2)+\cos(\alpha (y_1+y_2))-3]\ ,
\]
\[
\eta_3  =  \frac{2}{\al^3}[\sin(\al y_1)+\sin(\al y_2)-\sin(\al
(y_1 + y_2))]\ ,
\]
(cf. (\ref{v-Suth})) we get an algebraic form:
\[
h = h_{\rm Suth} + \la\,\eta_2 \equiv T_0 + \la\,T_1 \ ,
\]
where
\begin{eqnarray*}
h_{\rm Suth} &=&\hspace{-10pt}
        -(2{\eta}_2+\frac{\al^2}{2}{\eta}_2^2
        -\frac{\al^4}{24}{\eta}_3^2)
                \pa_{{\eta}_2{\eta}_2}^2
        -(6+\frac{4\al^2}{3}{\eta}_2)
        {\eta}_3\pa_{{\eta}_2{\eta}_3}^2\\
& \  & \hspace{-60pt}
        +(\frac{2}{3}{\eta}_2^2-\frac{\al^2}{2}{\eta}_3^2)
                \pa_{{\eta}_3{\eta}_3}^2
       \! +2(\nu+ \frac{1}{3})
(3 + \al^2\,\eta_2 )
                \pa_{{\eta}_2}
        \! +2(\nu+\frac{1}{3})\al^2{\eta}_3\pa_{{\eta}_3}
\end{eqnarray*}
\[
T_0 : {\cal P}_n^{(2)}(\eta) \mapsto {\cal P}_n^{(2)}(\eta), \quad
n\in \mathbb{N},
\]
\[
T_1 = \eta_2 \in {\cal P}_{1,2,3,\ldots}^{(2)} \ ,
\]

 Ground State: $\qquad \phi_0=1, \quad E_0=0$

{\it First correction:}

\[
 -\phi_1 = \frac{3}{2(1+3\nu)\al^2} \,\eta_2\ ,
\]
\[
 E_1= -\frac{3}{\al^2} \ .
\]
Since
\[
E_1 \ =\  \frac{\langle 0 \vert \eta_2(y)  \vert 0 \rangle
}{\langle 0 \vert 0 \rangle}
\]
we can find expectation value $\langle 0 \vert \eta_2(y)  \vert 0
\rangle$ {\it algebraically} using known normalization factor
$\langle 0 \vert 0 \rangle$ \cite{Olshanetsky:1983}.

{\it Second correction:}
\[
-\phi_2 = \frac{3}{8\al^4(1+3\nu)(1+6\nu)}\,
\bigg[(1+12\nu)\,\eta_2^2 +\frac{1}{4}\,\eta_3^2+ \frac{9 (2+13\nu
+12\nu^2)}{(1+3\nu)} \,\eta_2\bigg] \ ,
\]
\[
E_2= -\frac{27}{4\al^4}\, \frac{2+13\nu
+12\nu^2}{(1+3\nu)(1+6\nu)} \ .
\]

\section{\em CONCLUSION}

Algebraic forms of Calogero-Sutherland models give an opportunity
study their {\bf perturbations} by algebraic means through
developing a perturbation theory for single state.

Taking different perturbations and making comparison of present
perturbation theory with standard Rayleigh-Schroedinger
perturbation theory allow to calculate correlation functions for
Calogero-Sutherland models algebraically.

Algebraic forms of Calogero-Sutherland models allow to build their
Fock space representation (see \cite{Turbiner:1999}) and then
develop algebraic perturbation theory in Fock space. It gives a
chance to study isospectral discretizations of Calogero-Sutherland
models (on different lattices) and their perturbations
\cite{Turbiner:2001}.

\begingroup\raggedright
\endgroup

\end{document}